\pdfoutput=1
\documentclass[conference]{IEEEtran}
\usepackage{cite}
\usepackage{amsmath,amssymb,amsfonts}
\usepackage{graphicx}
\usepackage{textcomp}
\usepackage{booktabs}
\usepackage{xcolor}
\usepackage{url}
\usepackage{eso-pic}
\usepackage[hidelinks,
  pdftitle={PathGround: Auditing Existence, Gene Support, and Overclaim in LLM Pathway Interpretation},
  pdfauthor={Ye Ye}]{hyperref}
\newcommand{\doi}[1]{\href{https://doi.org/#1}{doi:~#1}}
\graphicspath{{figs/}}

\begin{document}

\title{PathGround: Auditing Existence, Gene Support,\\and Overclaim in LLM Pathway Interpretation}

\author{\IEEEauthorblockN{Ye Ye}
\IEEEauthorblockA{\textit{Independent Researcher}\\
New York, NY, USA\\
Email: yy4820@nyu.edu}}

\AddToShipoutPictureFG*{%
  \AtPageLowerLeft{%
    \put(\LenToUnit{0.5\paperwidth},\LenToUnit{0.32in}){%
      \makebox[0pt][c]{\parbox{\textwidth}{\centering\scriptsize
        Accepted to the TrustAI-BioD Workshop at the 2026 IEEE International Conference on Bioinformatics and Biomedicine (BIBM). DOI: to appear.\\
      \textcopyright~2026 IEEE. Personal use of this material is permitted. Permission from IEEE must be obtained for all other uses, in any current or future media, including reprinting/republishing this material for advertising or promotional purposes, creating new collective works, for resale or redistribution to servers or lists, or reuse of any copyrighted component of this work in other works.}}}}}

\maketitle

\begin{abstract}
Researchers increasingly paste a gene list into a large language model (LLM) and read back a few pathway names, each with a one-line justification. Such answers are usually scored by how similar the name is to a reference title. That score mixes three separate problems: the name may not exist in KEGG or Reactome, it may exist but not be supported by the input genes, or the justification may describe membership as activation or causation. We present PathGround, a deterministic, retrieval-free audit that checks each predicted pathway for existence, gene support, and overclaim wording, and re-queries the model after small edits to the gene list to measure stability. On 229 benchmark gene sets, GPT-4.1 named a real pathway in 96.7\% of claims, but only 70.8\% of claims were supported by the input genes. Unsupported claims were worded as cautiously as supported ones (93.8\% vs.\ 91.2\% conservative rationales). Randomly drawn database titles passed the existence check in every case yet were supported only 4.8\% of the time. The gap between existence and support persisted under all 48 support-threshold settings we tested. Shuffling the order of an unchanged gene list altered GPT-4.1's pathway set (mean Jaccard 0.68 on a 60-case subset). We therefore suggest reporting existence, support, overclaim, and stability separately when evaluating LLM pathway interpretation.
\end{abstract}


\section{Introduction}
Pathway interpretation is the step where a differential-expression table becomes a biological story. KEGG and Reactome encode that story as curated member lists~\cite{kanehisa2000kegg,milacic2024reactome}, and enrichment tools test whether a gene list overlaps those lists more than expected by chance~\cite{kuleshov2016enrichr}. LLMs offer a shortcut: give the model the genes and it returns pathway names with fluent explanations~\cite{hu2025evaluation,wang2025geneagent,zhu2025llm2geneset}. The shortcut saves time, but the user has little way of knowing which of the returned pathways the genes actually support. Models can invent pathway-like titles, attach real titles to genes that do not belong to them, and let a few well-known inflammatory or tumor-suppressor genes steer the explanation~\cite{hu2025evaluation,tiwari2023chatgpt,wu2025reactome}.

Most existing evaluations score the pathway name. Hu et al.\ compare LLM-generated process names with Gene Ontology titles by embedding similarity and then inspect rationales by hand~\cite{hu2025evaluation}, and llm2geneset scores generated gene sets against KEGG, Reactome, and WikiPathways~\cite{zhu2025llm2geneset}. Name similarity is useful, but it does not tell a fabricated pathway apart from a real one that is irrelevant to the input. An invented title is usually caught quickly, because a database search returns nothing. A real pathway with weak gene support is harder to notice: it passes a name lookup and often looks reasonable to a reader who does not check the member list. For biomedical use, the more relevant questions are whether an interpretation can be traced to evidence in the submitted genes, and whether it stays the same under changes that should not matter, such as the order in which the genes are listed.

PathGround is a simple, deterministic check for these questions. The model never sees the database while answering. After generation, we (i)~map each title onto KEGG or Reactome (\emph{existence}), (ii)~test whether enough input genes belong to the mapped pathway (\emph{support}), (iii)~flag rationales that describe membership as activation or causation (\emph{overclaim}), and (iv)~re-query after small edits to the gene list (\emph{stability}). We apply the audit to a 229-case benchmark built from public Enrichr libraries, which includes noisy, mixed-program, and decoy gene sets, and evaluate GPT-4.1 alongside three reference generators: random database titles, a closed-book recall baseline, and over-representation analysis (ORA). In our experiments GPT-4.1 rarely invented pathway names, but about one in four of its claims named a real pathway that the input genes did not support, and the wording of the rationale gave no hint of which claims these were.

The contribution is an evaluation protocol rather than a new interpretation system: (i)~a post-hoc, deterministic audit that separates three error types that name-similarity scores merge; (ii)~a reconstructible benchmark with noisy, mixed, and decoy splits whose ground truth is known by construction; and (iii)~reference generators that bound the metrics from below (random titles) and above (ORA).

\section{Related Work}
Several studies evaluate what LLMs say about gene sets. GSAI-style prompts ask a model for a process name, a rationale, and a confidence score; GPT-4 often recovers Gene Ontology titles on curated sets and lowers its confidence on random genes~\cite{hu2025evaluation}. llm2geneset uses LLMs to generate gene sets for over-representation analysis and compares them with curated libraries~\cite{zhu2025llm2geneset}. Azam et al.\ probe pathway knowledge in the opposite direction, asking models to list KEGG members and gene relations for a given pathway~\cite{azam2024comprehensive}. Reactome curators have tested LLMs for gap filling and gene annotation and report useful suggestions mixed with incorrect gene--pathway links and unreliable references~\cite{tiwari2023chatgpt,wu2025reactome}. In these evaluations the unit is usually a name or a gene list, so a real but irrelevant KEGG map and an invented ``neuroinflammatory response pathway'' can receive similar scores if both mention inflammation.

A second line of work changes the generator. GeneAgent checks its own claims against g:Profiler and other databases~\cite{wang2025geneagent}. React-to-Me answers questions with retrieval over Reactome~\cite{mohammadi2025reacttome}. cGSA and MAPA place an LLM after enrichment analysis, where it filters, clusters, or summarizes significant pathways~\cite{wang2025cgsa,ge2025mapa}. PathGround is complementary to these systems and differs from them in role. GeneAgent and retrieval-based assistants are \emph{generators}: they consult databases while answering, and their internal checks are part of the method. PathGround is an external \emph{evaluator}: it never changes the answer, applies the same library and rules to every system, and reports existence, support, overclaim, and stability separately. It can therefore be applied to a plain chat model, which is what most users have, to systems that do not expose a retrieval trace, and to GeneAgent-style systems themselves, to test whether their grounding improves support or only vocabulary (Section~\ref{sec:discussion}).

\section{The PathGround Audit}
\subsection{Task}
A system receives only a human gene set $G=\{g_1,\ldots,g_n\}$ and returns $k\in[3,5]$ claims $\{(p_i,r_i)\}$, where $p_i$ is a pathway title and $r_i$ is a one-sentence rationale. No database, tool, or document index is provided.

\subsection{Level 1: Existence}
The reference library $\mathcal{L}$ is the union of the Enrichr KEGG 2021 Human and Reactome 2022 libraries~\cite{kuleshov2016enrichr}, restricted to sets with at least eight genes (314 KEGG and 1{,}593 Reactome sets; 12{,}120 unique symbols). Titles are normalized by lowercasing, removing identifiers such as \texttt{R-HSA-*} and \texttt{hsa*}, replacing punctuation with spaces, and applying a short alias list (e.g., PI3K-Akt, NF-$\kappa$B/NF-kB, TCA/Krebs, JAK-STAT, TGF-$\beta$). Generic words such as \emph{pathway}, \emph{signaling}, and \emph{human} are dropped before token comparison. A predicted title $p$ maps to library entry $\ell$ if its match score $s(p,\ell)$ is at least $0.62$, where
\[
s(p,\ell)=\begin{cases}
1 & \text{equal normalized titles},\\
0.92 & \text{containment and } J\ge 0.4,\\
0.55J+0.45O & \text{otherwise}.
\end{cases}
\]
Here $J$ is the token Jaccard index, $O$ is token overlap relative to the shorter title, and containment means one normalized title is a substring of the other. Titles that map to nothing are labeled \emph{not in library}. We report an exact-match ablation in Section~\ref{sec:ablation}.

Reactome is a hierarchy in which a parent event (e.g., \emph{Cell Cycle}) contains child events with smaller member sets, whereas KEGG maps are flat, and the two databases often name the same biology differently. The matcher does not use the hierarchy: a claim naming a parent maps to the parent entry, and a near-synonym absent from the alias list is labeled not in library. Because Level~2 uses the member set of whichever entry was matched, name matching decides only which member list is tested; support is decided by the genes.

\subsection{Level 2: Gene support}
If $p$ maps to a pathway with member set $P$, the claim is \emph{supported} when
\[
\begin{aligned}
&|G\cap P|\ge 3 \quad\text{and}\\
&\Pr[X\ge|G\cap P|]\le 0.05 \;\;\text{or}\;\; |G\cap P|/|G|\ge 0.08,
\end{aligned}
\]
where $X$ follows a hypergeometric distribution with population $N=12{,}120$, $|P|$ successes, and $|G|$ draws. Otherwise the claim is \emph{real but unsupported}: the pathway exists, but the input does not justify it. A name lookup cannot detect this kind of error.

The thresholds are design choices, fixed before scoring. The minimum of three overlapping genes prevents a single hub gene, or a pair of genes shared by many maps, from justifying a pathway. The hypergeometric criterion uses the conventional 0.05 level against the library background. The 8\% overlap alternative covers small or focused gene sets, for which a restricted background makes the hypergeometric test conservative. All three rules are applied identically to every generator. Section~\ref{sec:ablation} varies each of them and shows that they shift absolute support rates but not the gap between existence and support.

\emph{Example.} In one decoy case the input contains 15 TCA-cycle genes plus the interferon-pathway genes IFNA8, STAT1, and SOCS5. GPT-4.1 names Reactome's \emph{Interferon Signaling}. The title exists in the library, so the claim passes Level~1. Only two input genes are members, so it fails Level~2 and is counted as real but unsupported. A name-similarity score would treat it as a correct answer.

\subsection{Level 3: Overclaim}
Gene membership alone does not mean that a pathway is activated. A rationale is tagged \emph{overclaim} if it states that the pathway is activated, inhibited, up- or down-regulated, induced, driven, triggered, or causally responsible, and it does not also use membership language (\emph{involved in}, \emph{participate in}, \emph{components of}, \emph{consistent with}, \emph{annotated in KEGG/Reactome}). Sentences that contain both are tagged \emph{mixed}. The overclaim rate is the fraction tagged overclaim or mixed. This level reads $r_i$ only and needs no additional model call. Because it is a lexicon, it detects explicit activation and causal verbs but not subtler mechanistic framing, such as ``plays a central role in,'' ``suggests that the cell relies on,'' or a hedged causal statement without a listed verb. Its rate should therefore be read as a lower bound on overclaim.

\subsection{Stability}
For a case with normalized pathway set $P_0$, we build $G'$ by shuffling the order of $G$, dropping 10\% or 20\% of its genes, or adding 10\% random library genes. We re-query the generator and report $\mathrm{Jaccard}(P_0,P')$. A set-based interpretation should be unaffected by shuffling, and small edits should cause only small changes.

\subsection{Metrics}
All rates are case means. \textbf{Existence} is the fraction of claims that map to $\mathcal{L}$. \textbf{Supported} and \textbf{unsupported} split all claims by Level~2, and \textbf{not in library} is the remainder. \textbf{Recovery} is the fraction of planted pathways matched by at least one claim. \textbf{Decoy preference} is 1 when claims mapped to the planted decoy outnumber supported claims for the majority pathway. \textbf{Overclaim} and \textbf{stability} are defined above.

\section{Benchmark and Generators}
\subsection{Benchmark}
All gene sets are sampled from $\mathcal{L}$ with a fixed seed, so the benchmark is fully determined by the two public library files. There are five splits.
\emph{Clean} (162): 10, 20, or 30 members of one of 30 KEGG or 25 Reactome seed pathways.
\emph{Noisy} (33): 14 members plus 6 genes from outside the pathway ($\approx$30\% contamination).
\emph{Mixed} (14): 12 genes each from two disjoint pathways (e.g., apoptosis and cell cycle, OXPHOS and glycolysis); a complete answer names both.
\emph{Decoy} (14): 15 genes unique to a majority pathway plus 5 genes unique to a salient ``foil'' (TNF, NF-$\kappa$B, p53, JAK-STAT, TLR, IL-17, chemokine signaling, and similar).
\emph{GEO-like} (6): eight genes from each of three related KEGG maps plus eight random genes, meant to approximate a differential-expression list. Of the 229 cases, 133 are seeded from KEGG and 96 from Reactome.

\subsection{Generators}
Four generators are scored by the same checker.

\textbf{Random titles.} Four titles drawn uniformly from $\mathcal{L}$. They exist by construction, so this control shows what a name-only check would reward.

\textbf{Closed-book baseline.} A deterministic marker-recall model over 39 common pathway titles and a small dictionary of well-known genes. When evidence is weak it emits a plausible umbrella title (e.g., ``metabolic reprogramming pathway''), and with probability 0.22 it appends TNF signaling whenever a \texttt{MAPK*}, \texttt{TNF*}, \texttt{IL*}, or \texttt{STAT*} gene is present. It is a transparent stand-in for ungrounded recall, not a model of any specific LLM.

\textbf{GPT-4.1.} All 229 cases were sent to the OpenAI Chat Completions API (\texttt{gpt-4.1-2025-04-14}, temperature 0, no tools, no retrieval). The prompt asks for 3--5 KEGG or Reactome pathways, each with a one-sentence reason, in JSON, and states that a pathway should be claimed only if multiple input genes participate in it. Six replies with malformed JSON were re-requested with a shorter output format. The run used 58.4k prompt and 64.7k completion tokens.

\textbf{ORA.} A one-sided hypergeometric test of $G$ against every entry in $\mathcal{L}$; the five lowest $p$-values are returned. ORA reads the library at prediction time and serves as a grounded upper reference.

The generators play different roles; the comparison is not a contest between GPT-4.1 and ORA. ORA reads the library, so it should score near the ceiling and checks that the audit does not penalize grounded answers. Random titles show what a name-only check rewards, and the closed-book baseline is a transparent ungrounded control. The quantity of interest for GPT-4.1 is the gap between its own existence and support rates.

\subsection{Scope}
The benchmark is meant to test whether the audit separates these error types, not to rank models. All gene sets are synthetic but reconstructible from the library files, and only one LLM is evaluated in full. Synthetic sets give known ground truth, which real data do not, but they are cleaner than real differential-expression lists: those contain co-regulated genes from many programs, cross-talk between pathways, and genes absent from both databases, and their true pathways are unknown. The six GEO-like cases are an approximation of this setting, not a substitute for it, and results on them are illustrative only.

\section{Results}
\subsection{Existence and gene support}
\begin{figure}[t]
\centering
\includegraphics[width=\columnwidth]{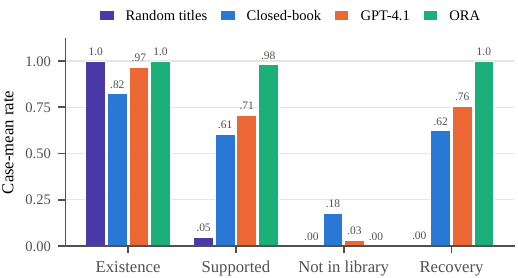}
\caption{Existence, support, not-in-library, and recovery rates for four generators scored against the same library.}
\label{fig:overall}
\end{figure}

Table~\ref{tab:main} and Fig.~\ref{fig:overall} summarize the four generators. Random titles pass the existence check in every case, yet only 4.8\% of them are supported and they recover almost none of the planted pathways. A check that only asks whether a title appears in KEGG or Reactome would give these random guesses a perfect score.

GPT-4.1 rarely invents names. Only 30 of its 1{,}019 claims (3.3\% case mean) do not map to the library, and inspection shows that several of these are paraphrases of real entries, such as the spelled-out form of KEGG's PPAR signaling pathway, so 3.3\% is an upper bound on outright fabrication. Support, however, is only 70.8\%: roughly one claim in four names a real pathway that the input genes do not justify. The closed-book baseline shows the same gap at a lower level (82.2\% existence, 60.5\% support), while ORA reaches 97.9\% support and recovers every planted pathway. The remaining 2.1\% of ORA claims are lower-ranked hits that fail the overlap rule and are counted as unsupported like any other claim.

\begin{table}[t]
\centering
\caption{Case-mean rates. Unsupp.\ = real but unsupported; Not in lib.\ = does not map to KEGG or Reactome.}
\label{tab:main}
\setlength{\tabcolsep}{3pt}
\footnotesize
\begin{tabular}{llrrrrrr}
\toprule
Generator & Split & $n$ & Exist. & Supp. & Unsupp. & Not in lib. & Recov. \\
\midrule
Random & All & 229 & 1.000 & .048 & .952 & .000 & .004 \\
Closed-book & All & 229 & .822 & .605 & .217 & .178 & .623 \\
\midrule
GPT-4.1 & All & 229 & .967 & .708 & .259 & .033 & .755 \\
GPT-4.1 & Clean & 162 & .969 & .721 & .248 & .031 & .728 \\
GPT-4.1 & Noisy & 33 & .971 & .632 & .339 & .029 & .758 \\
GPT-4.1 & Mixed & 14 & .939 & .736 & .204 & .061 & .857 \\
GPT-4.1 & Decoy & 14 & .950 & .643 & .307 & .050 & .929 \\
GPT-4.1 & GEO-like & 6 & 1.000 & .867 & .133 & .000 & .833 \\
\midrule
ORA & All & 229 & 1.000 & .979 & .021 & .000 & 1.000 \\
\bottomrule
\end{tabular}
\end{table}

\subsection{Sources of unsupported claims}
Unsupported GPT-4.1 claims are mostly familiar, high-degree pathways rather than obscure ones. The most frequent are Ubiquitin mediated proteolysis (23 claims), MAPK signaling (12), Antigen processing and presentation (11), p53 signaling (9), Apoptosis (8), PI3K-Akt signaling (7), and the Reactome Cell Cycle parent (7). These maps share kinases, ubiquitin ligases, and immune genes with many other programs, so a few input genes are enough to make them sound plausible. The closed-book baseline shows the same pattern (MAPK, p53, NF-$\kappa$B, and TNF signaling top its list), while its titles that are not in the library are umbrella phrases such as ``oncogenic kinase cascade'' and ``stemness maintenance.''

\begin{figure}[t]
\centering
\includegraphics[width=\columnwidth]{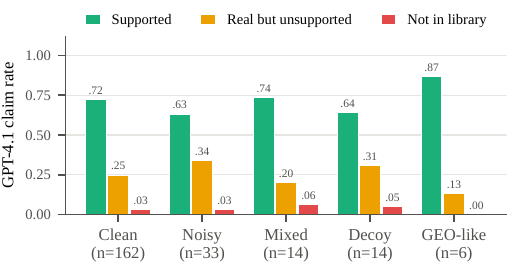}
\caption{GPT-4.1 claims by benchmark split: supported, real but unsupported, and not in library.}
\label{fig:split}
\end{figure}

The mix of errors shifts across splits (Fig.~\ref{fig:split}). Contamination lowers GPT-4.1 support from 72.1\% on clean sets to 63.2\% on noisy sets, and decoy genes push the unsupported share to 30.7\%. GEO-like unions are the easiest split (86.7\% support), because several related pathways are sampled together and well-known markers are dense. Results on such cases alone would overstate performance.

\subsection{KEGG- versus Reactome-origin cases}
\begin{figure}[t]
\centering
\includegraphics[width=\columnwidth]{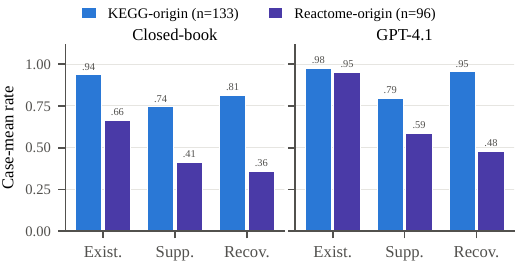}
\caption{Existence, support, and recovery by source of the planted gene set for the closed-book baseline and GPT-4.1.}
\label{fig:source}
\end{figure}

Performance depends on which database the planted pathway comes from (Fig.~\ref{fig:source}). For GPT-4.1, existence is similar for KEGG- and Reactome-origin cases (97.7\% vs.\ 95.3\%), but support drops from 79.4\% to 58.8\% and recovery from 95.5\% to 47.9\%. The model tends to answer with short KEGG-style names (Apoptosis, Wnt signaling, JAK-STAT) even when the genes come from a longer, hierarchical Reactome entry such as \emph{Cytokine Signaling in Immune system}. The resulting KEGG neighbors exist but only partly overlap the Reactome members. The closed-book baseline shows the same asymmetry more sharply (existence 93.7\% vs.\ 66.4\%). ORA recovers every planted pathway from both sources.

Three differences between the databases contribute to the gap. First, naming: KEGG titles are short and widely used in the literature, whereas Reactome titles are longer, event-style descriptions that a model rarely reproduces verbatim, so recovery, which requires the planted title, falls sharply. Second, structure: a Reactome entry can sit inside a hierarchy of broader and narrower events, and a KEGG map or a Reactome parent with a related name has a different member set, so the claim exists but meets the overlap rule less often. Third, granularity: Reactome divides processes into many specific events, so even a closely related entry may share fewer than three genes with the input. The gap therefore reflects how the model names biology as much as what it knows, which is why we score support on gene membership rather than on title identity.

\subsection{Effect of decoy genes}
Each decoy case adds five genes from a well-known foil pathway to fifteen genes from a less prominent majority pathway. Because the foil genes are genuine signal, naming the foil can be correct; we count it as a problem when it outranks the majority pathway or pulls in unsupported neighbors. The closed-book baseline names the foil in 6 of 14 cases and prefers it in 2. GPT-4.1 is more restrained: it names the exact foil in 2 of 14 cases, prefers it in 1, and recovers the majority pathway in 13. The foil genes still affect its answers (Table~\ref{tab:decoy}). With a TCA-cycle majority plus IFNA8, STAT1, and SOCS5, GPT-4.1 adds two interferon pathways that each overlap only two input genes. With a Notch majority plus IL-17 pathway genes, it adds cytokine--cytokine receptor interaction and an antigen-presentation pathway with no overlap at all. A name-only check would accept all of these titles.

\begin{table}[t]
\centering
\caption{Selected GPT-4.1 decoy cases. Supp.\ = supported / emitted claims. Unsupported claims are listed with their overlap $|G\cap P|$.}
\label{tab:decoy}
\setlength{\tabcolsep}{2.5pt}
\footnotesize
\begin{tabular}{@{}p{1.25cm}p{1.2cm}cp{4.2cm}@{}}
\toprule
Majority & Foil & Supp. & Unsupported claims (overlap) \\
\midrule
TCA cycle & JAK-STAT & 1/5 & Interferon Signaling (2); Regulation of IFNA signaling (2); Gluconeogenesis (2); EGF/EGFR signaling (1) \\
Glycolysis & TLR & 2/5 & Pyruvate metabolism (2); Interferon Signaling (0) \\
Notch & IL-17 & 2/5 & Cytokine--cytokine receptor interaction (2); Class I MHC antigen presentation (0) \\
Ferroptosis & TCR & 3/5 & MAPK signaling (2); PI3K-Akt signaling (1) \\
Hippo & TNF & 3/5 & TGF-$\beta$ signaling (2); MAPK signaling (1); majority not recovered \\
Cell cycle & p53 & 4/5 & Apoptosis (2); foil named and preferred \\
\bottomrule
\end{tabular}
\end{table}

\subsection{Rationale wording}
Our prompt tells the model to claim a pathway only when several input genes participate, and GPT-4.1 largely complies in its wording. Of 1{,}019 rationales, 91.4\% use membership language only (``involved in,'' ``participate in,'' ``components of''), 7.4\% are mixed, and 1.3\% make a strict activation or causal claim. Overclaim is therefore rare in this run. Unsupported claims are also worded as cautiously as supported ones: 93.8\% of unsupported rationales are conservative, compared with 91.2\% of supported rationales, so the tone of a rationale says little about whether the claim is supported. A language-only hallucination check would miss most of the errors in this run, which is why we score support and wording separately.

\subsection{Stability under perturbation}
\label{sec:stability}
\begin{figure}[t]
\centering
\includegraphics[width=\columnwidth]{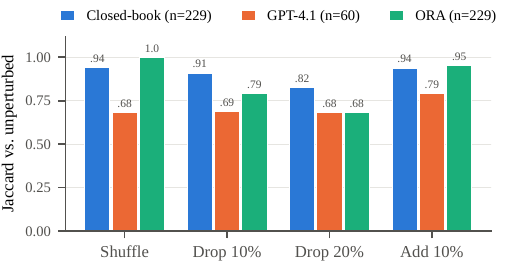}
\caption{Mean Jaccard between the pathway sets returned before and after each perturbation.}
\label{fig:stab}
\end{figure}

Table~\ref{tab:stab} and Fig.~\ref{fig:stab} report the Jaccard overlap between pathway sets before and after editing $G$. ORA does not depend on gene order (Jaccard 1.00), and the closed-book baseline changes only through its random TNF rule (0.94). GPT-4.1 is less stable. Because each perturbation requires new API calls, we re-queried 60 cases rather than all 229: every decoy (14), mixed (14), and GEO-like (6) case, plus the first 10 noisy and the first 16 clean cases in benchmark order. The subset therefore over-represents the harder splits and is mostly KEGG-origin (54 of 60). Re-querying with the same genes in shuffled order yields a mean Jaccard of 0.68, and only 27\% of shuffled re-queries return the same pathway set. The effect appears in every split (clean 0.66, noisy 0.71, mixed 0.76, decoy 0.61, GEO-like 0.65), so it is not confined to the decoy cases, although it is largest there. Dropping 10\% or 20\% of the genes produces about the same change (0.69 and 0.68), so reordering alone moves GPT-4.1's answer as much as deleting a fifth of the input. Adding 10\% random genes has a smaller effect (0.79). In typical shuffled runs the model keeps the majority pathway but swaps a secondary one. On two decoy cases the swap brings in the planted foil itself: p53 signaling is replaced by TNF signaling in the apoptosis case, and cAMP signaling by NF-$\kappa$B signaling in the Wnt case. Since temperature was 0, this points to sensitivity to input order rather than sampling noise, although API-side nondeterminism cannot be fully excluded.

\begin{table}[t]
\centering
\caption{Mean Jaccard of normalized pathway sets after perturbation. Closed-book and ORA: $n=229$; GPT-4.1: 60-case subset.}
\label{tab:stab}
\footnotesize
\begin{tabular}{lccc}
\toprule
Perturbation & Closed-book & GPT-4.1 & ORA \\
\midrule
Shuffle order & .940 & .680 & 1.000 \\
Drop 10\% of genes & .909 & .686 & .792 \\
Drop 20\% of genes & .824 & .680 & .682 \\
Add 10\% random genes & .939 & .790 & .951 \\
\bottomrule
\end{tabular}
\end{table}

\subsection{Sensitivity to matching and support thresholds}
\label{sec:ablation}
Requiring exact normalized matches instead of fuzzy matching lowers closed-book existence from 82.2\% to 72.2\% and recovery from 62.3\% to 55.2\%. Support stays well below existence under both rules (53.5\% exact, 60.5\% fuzzy). The alias list mainly helps with KEGG/Reactome wording differences; it does not turn an unsupported claim into a supported one.

We also re-scored the stored outputs of all four generators while varying each support parameter: the minimum overlap (2--5), the $p$-value cutoff (0.01--0.10, or no $p$ criterion), the overlap-fraction alternative (0.05--0.12, or none), and the title-match cutoff (0.55--0.80, or exact). Table~\ref{tab:thresh} lists representative settings. The minimum overlap is the only parameter with a large effect: GPT-4.1 support ranges from 84.6\% with two genes to 49.9\% with five, because many unsupported claims share exactly two genes with the input. The $p$-value and fraction criteria barely matter, since three or more overlapping genes are almost always significant against this background. Across all 48 combinations of overlap, $p$, and fraction settings, GPT-4.1 support stays between 49.5\% and 85.4\% while its existence is 96.7\%, random titles never exceed 8.0\% support, and ORA never falls below 89.0\%. KEGG-origin cases are better supported than Reactome-origin cases under every setting. The thresholds change absolute support, not the conclusions.

\begin{table}[t]
\centering
\caption{Support under alternative thresholds (case means; other parameters at default). Existence for GPT-4.1 depends only on the match rule.}
\label{tab:thresh}
\setlength{\tabcolsep}{3pt}
\footnotesize
\begin{tabular}{lcccc}
\toprule
 & \multicolumn{2}{c}{GPT-4.1} & Random & ORA \\
\cmidrule(lr){2-3}
Setting & Exist. & Supp. & Supp. & Supp. \\
\midrule
Default ($\ge$3 genes, $p\le$0.05 or $\ge$8\%) & .967 & .708 & .048 & .979 \\
Min.\ overlap 2 & .967 & .846 & .074 & .995 \\
Min.\ overlap 4 & .967 & .604 & .028 & .947 \\
Min.\ overlap 5 & .967 & .499 & .020 & .890 \\
$p\le$0.01 & .967 & .708 & .048 & .979 \\
$p$ criterion only (no fraction) & .967 & .702 & .046 & .979 \\
Match cutoff 0.70 & .962 & .707 & .048 & .979 \\
Exact title match & .831 & .611 & .045 & .938 \\
\bottomrule
\end{tabular}
\end{table}

\section{Discussion}
\label{sec:discussion}
\textbf{Implications for trustworthy use.} The four checks answer different questions: whether the pathway exists, whether the input genes belong to it, whether the rationale claims more than membership, and whether the answer changes when the input changes in ways that should not matter. In our results they disagree. Random titles exist but are rarely supported (100\% vs.\ 5\%). GPT-4.1 fails the support check in 29\% of its claims, while under 9\% of its rationales carry any overclaim tag. The shuffle test shows that setting temperature to 0 does not guarantee the same answer for the same genes. Because the checks run after generation and need no access to the model, they could also be used to flag unsupported claims in an interface before a user acts on them.

\textbf{Evaluating grounded systems.} Retrieval-augmented and self-verifying systems~\cite{wang2025geneagent,mohammadi2025reacttome} should be evaluated with the same metrics rather than with another similarity score. If grounding works, unsupported and not-in-library rates should both fall, and decoy preference should fall on the decoy split. If grounding only constrains the vocabulary, existence will rise while support stays flat.

\textbf{Limitations.} We evaluate one commercial model with one prompt, so the GPT-4.1 rates should not be generalized to other models. The closed-book baseline shows the same existence--support gap, which suggests the gap is not specific to GPT-4.1, but it is a transparent stand-in, not an LLM; testing several mainstream LLMs is the most important next step. The stability analysis uses a 60-case subset that over-represents the harder splits and contains only six Reactome-origin cases, and its per-split estimates rest on 6 to 16 cases each. Fuzzy matching can credit a parent pathway when the model names a child, or miss a synonym absent from the alias list; using the Reactome event hierarchy and curated synonym lists would make the existence check more systematic. The overlap thresholds and the library-restricted background are design choices; the minimum overlap in particular changes absolute support rates (Table~\ref{tab:thresh}), so support percentages should be reported together with the rule used. The benchmark is synthetic, and the six GEO-like cases do not represent real expression data. Level~3 uses a lexicon, scores only the rationale sentence, and misses implicit mechanistic claims; an entailment model or expert annotation would be needed to measure such claims.

\section{Conclusion}
PathGround checks LLM pathway interpretations against KEGG and Reactome after generation. On 229 gene sets, GPT-4.1 named a real pathway in 96.7\% of claims, but only 70.8\% were supported by the input genes; unsupported claims were worded as cautiously as supported ones; and shuffling the gene order changed the returned pathway set (mean Jaccard 0.68). Reporting existence, support, overclaim, and stability separately makes these problems visible, and the same checks can be applied without change to other models or to retrieval-augmented systems.

\section*{Acknowledgment}
Generative AI tools (xAI Grok, OpenAI ChatGPT, and Anthropic Claude) were used for portions of code implementation, literature discovery, and language and editorial revision. The author developed the research idea, study design, evaluation protocol, analysis, and interpretation, and reviewed and verified all AI-assisted outputs.

\section*{Data and code}
The benchmark is generated with a fixed seed from the public Enrichr KEGG 2021 Human and Reactome 2022 libraries. The benchmark, the PathGround checker, all model outputs, and the analysis scripts are available at \url{https://github.com/biubiutomato/pathground-llm-audit}.


\begin{thebibliography}{99}
\bibitem{kanehisa2000kegg}
M.~Kanehisa and S.~Goto, ``KEGG: Kyoto encyclopedia of genes and genomes,'' \emph{Nucleic Acids Res.}, vol.~28, no.~1, pp.~27--30, 2000, \doi{10.1093/nar/28.1.27}.

\bibitem{milacic2024reactome}
M.~Milacic \emph{et al.}, ``The Reactome pathway knowledgebase 2024,'' \emph{Nucleic Acids Res.}, vol.~52, no.~D1, pp.~D672--D678, 2024, \doi{10.1093/nar/gkad1025}.

\bibitem{kuleshov2016enrichr}
M.~V. Kuleshov \emph{et al.}, ``Enrichr: A comprehensive gene set enrichment analysis web server 2016 update,'' \emph{Nucleic Acids Res.}, vol.~44, no.~W1, pp.~W90--W97, 2016, \doi{10.1093/nar/gkw377}.

\bibitem{hu2025evaluation}
M.~Hu \emph{et al.}, ``Evaluation of large language models for discovery of gene set function,'' \emph{Nat. Methods}, vol.~22, pp.~82--91, 2025, \doi{10.1038/s41592-024-02525-x}.

\bibitem{wang2025geneagent}
Z.~Wang \emph{et al.}, ``GeneAgent: Self-verification language agent for gene-set analysis using domain databases,'' \emph{Nat. Methods}, vol.~22, no.~8, pp.~1677--1685, 2025, \doi{10.1038/s41592-025-02748-6}.

\bibitem{zhu2025llm2geneset}
J.~Zhu, R.~Y. Wang, X.~Wang \emph{et al.}, ``Enhancing gene set overrepresentation analysis with large language models,'' \emph{Bioinform. Adv.}, vol.~5, no.~1, Art.~no.~vbaf054, 2025, \doi{10.1093/bioadv/vbaf054}.

\bibitem{azam2024comprehensive}
M.~Azam \emph{et al.}, ``A comprehensive evaluation of large language models in mining gene relations and pathway knowledge,'' \emph{Quant. Biol.}, vol.~12, no.~4, pp.~360--374, 2024, \doi{10.1002/qub2.57}.

\bibitem{tiwari2023chatgpt}
K.~Tiwari \emph{et al.}, ``ChatGPT usage in the Reactome curation process,'' \emph{bioRxiv}, 2023, \doi{10.1101/2023.11.08.566195}.

\bibitem{wu2025reactome}
G.~Wu \emph{et al.}, ``Application of large language models for annotating genes into Reactome pathways,'' \emph{bioRxiv}, 2025, \doi{10.64898/2025.12.20.695723}.

\bibitem{wang2025cgsa}
Z.~Wang \emph{et al.}, ``Knowledge-guided contextual gene set analysis with large language models,'' \emph{Bioinformatics}, vol.~42, suppl.~1, Art.~no.~btag214, 2026, \doi{10.1093/bioinformatics/btag214}.

\bibitem{mohammadi2025reacttome}
H.~Mohammadi \emph{et al.}, ``React-to-Me: A conversational interface for interactive exploration of the Reactome pathway knowledgebase,'' \emph{bioRxiv}, 2025, \doi{10.64898/2025.12.12.693752}.

\bibitem{ge2025mapa}
Y.~Ge \emph{et al.}, ``Leveraging large language models for redundancy-aware pathway analysis and deep biological interpretation,'' \emph{bioRxiv}, 2025, \doi{10.1101/2025.08.23.671949}.
\end{thebibliography}
\end{document}